\documentclass[12pt,preprint]{aastex}

\shorttitle{Explosion Driven by BH Accretion in SNe IIP}
\shortauthors{Utrobin, Chugai, and Botticella}

\begin{document}

\title{Type IIP Supernova 2009kf: Explosion Driven by Black Hole Accretion?}

\author{
V. P. Utrobin\altaffilmark{1,2},
N. N. Chugai\altaffilmark{3},
and
M. T. Botticella\altaffilmark{4}
}
\altaffiltext{1}{Institute of Theoretical and Experimental Physics,
   B.~Cheremushkinskaya St. 25, 117218 Moscow, Russia; utrobin@itep.ru}
\altaffiltext{2}{Max-Planck-Institut f\"ur Astrophysik,
   Karl-Schwarzschild-Str. 1, D-85741 Garching, Germany}
\altaffiltext{3}{Institute of Astronomy, RAS, Pyatnitskaya 48, 109017 Moscow, Russia;
   nchugai@inasan.ru}
\altaffiltext{4}{Astrophysics Research Centre, School of Maths and Physics,
   Queen’s University, BT7 1NN, Belfast, UK; m.botticella@qub.ac.uk}

\begin{abstract}

Unusually bright type IIP supernova (SN) 2009kf is studied employing the hydrodynamic
   modelling.
We derived optimal values of the ejecta mass of $28.1~M_{\sun}$, explosion energy of
   $2.2\times10^{52}$ erg, and presupernova radius of $2\times10^3~R_{\sun}$ assuming
   that $^{56}$Ni mass is equal to the upper limit of $0.4~M_{\sun}$.
We analyzed effects of the uncertainties in the extinction and $^{56}$Ni mass and
   concluded that both the ejecta mass and explosion energy cannot be significantly
   reduced compared with the optimal values.
The huge explosion energy of SN~2009kf indicates that the explosion is caused by
   the same mechanism which operates in energetic SNe~Ibc (hypernovae), i.e., via
   a rapid disk accretion onto black hole.
The ejecta mass combined with the black hole mass and the mass lost by stellar wind yields
   the progenitor mass of about $36~M_{\sun}$.
We propose a scenario in which massive binary evolution might result in the SN 2009kf
   event.

\end{abstract}


\section{Introduction}
\label{sec-intro}

Type IIP supernovae (SNe~IIP) are believed to originate from $9-25~M_{\sun}$ stars
   which end their life with the core collapse into neutron star and the subsequent
   envelope ejection via the neutrino-driven mechanism \citep{HFWLH_03}, or
   alternatively, via the magneto-rotational mechanism  \citep{MBA_06}.
In fact, limits for the mass range of SNe~IIP progenitors (main-sequence stars) are
   fuzzy, because the theory of both stellar evolution and SN~IIP explosion is still
   in progress and cannot predict reliably the outcome.
On the other hand, the observational constraints for the progenitor masses are rather
   ambiguous.
Indeed, the progenitors of SNe~IIP recovered from pre-explosion images turn out to be
   predominantly low-mass ($8-17~M_{\sun}$) stars \citep{Sma_09}, while the hydrodynamic
   modelling of the handful of well-observed SNe~IIP indicates that SNe~IIP primarily
   originate from high-mass ($15-25~M_{\sun}$) progenitors
   \citep[][and references there]{UC_09}.

More massive ($25-100~M_{\sun}$) stars end up with the black hole formation which may
   be accompanied with a very weak explosion \citep{HFWLH_03}; they are associated with
   underluminous SNe~IIP \citep{Tur_98}. 
Alternatively, the underluminous SNe~IIP might originate from the low-mass end of massive
   star range \citep{CU_00,KJH_06,UCP_07,UC_08}.
A tiny fraction (order of 1\%) of stars producing black holes is responsible for
   energetic SNe (hypernovae) powered by the disk accretion onto black hole
   \citep{Gal_98,Iwa_98,MW_99,MWH_01,HFWLH_03}.
It should be stressed that all the hypernovae so far observed are exploding Wolf-Rayet
   (WR) stars lacking any hydrogen in the outer layers.

Recently a distant ($z=0.182$) type IIP SN~2009kf has been discovered by the Pan-STARRS 1
   survey \citep{You_09} which turns out to be unusually luminous for this class of SNe.
With the mid-plateau magnitude $M_V\approx -18.4$ mag it is $1.5-2$ mag brighter compared
   to normal SNe~IIP.
Despite a large distance this SN became a subject of the detailed photometric and
   spectroscopic study \citep{Bot_10}.
Apart from the high optical luminosity the SN is unusually bright in the ultraviolet (UV)
   band of {\em GALEX} with the near UV absolute magnitude $M_{\rm NUV}=-21.5$ mag and has
   unusually large photospheric velocity $\sim9000$ km\,s$^{-1}$ on day 61 \citep{Bot_10}.
Authors suggest that the standard model of SN~IIP is not applicable to SN~2009kf.
They mention several possibilities to account for this unusual SN~IIP: a huge explosion
   energy ($>10^{52}$ erg), a large pre-SN radius ($>1000 R_{\sun}$), a large $^{56}$Ni
   mass, and a strong interaction of ejecta with a dense circumstellar (CS) shell.

Given unusual characteristics of SN~2009kf we present results of the hydrodynamic
   simulations of this object in a framework of the explosion of a red supergiant (RSG).
We recover basic parameters of the event and find that the required explosion energy
   is indeed tremendous, comparable with the explosion energy of hypernovae.
We evaluate the role of alternative power sources, viz. radioactive $^{56}$Ni decay and
   CS interaction, and find them irrelevant.
Implications of our results for the pre-SN and the progenitor are discussed.

\section{Hydrodynamic Model and Presupernova}
\label{sec-mod}

The spherically-symmetric hydrodynamic code with one-group radiation transfer
   is employed to model SN~2009kf.
The code was described in details earlier \citep{Utr_04} and was used to study several
   other well-observed SNe~IIP \citep[][and references there]{UC_09}.
The explosion energy is modelled by the supersonic piston applied at the mass-cut
   which presumably is a border between collapsing core and ejected mass.
A non-evolutionary massive RSG in hydrostatic equilibrium is used as a pre-SN model.
The helium core of pre-SN is presumably mixed with the hydrogen envelope and the density
   jump between the helium core and hydrogen envelope is essentially smoothed.
Arguments in favor of a non-evolutionary pre-SN for the hydrodynamic models of SNe~IIP
   were presented by \citet{UC_08}.

Principal fitting parameters, which determine the light curve and the expansion
   velocity at the photosphere, are the explosion energy, ejecta mass, pre-SN radius,
   and amount of $^{56}$Ni.
"Second order" but also important ingredient of the model is a mixing between helium core
   and hydrogen envelope.
The mixing determines the shape of the plateau at the final stage, and generally a high
   degree of mixing is needed to account for the light curve shape \citep{Utr_07}.
A mixing of $^{56}$Ni also affects the light curve at the transition from the plateau to
   the radioactive tail.
The issue of the dependence of the hydrodynamic model on parameters was explored in
   detail earlier \citep{Utr_07}.

Preliminary computations of SN~2009kf light curves and expansion velocities showed that
   the explosion energy in the optimal model should exceed $10^{52}$ erg.
That enormous energy is beyond the capabilities of the explosion mechanisms usually
   associated with the core collapse into the neutron star and seems to indicate
   the same explosion mechanism as for hypernovae \citep{MW_99,MWH_01,HFWLH_03}.
We therefore consider below models with the large mass of collapsing core that forms
   presumably a black hole with the rest mass of $4.5~M_{\sun}$.
The latter is close to the minimal value which provides the required explosion energy
   by the accretion of $\sim 2~M_{\sun}$ for $\sim$ 1\% efficiency.

\section{Results}
\label{sec-res}

\begin{figure}
\epsscale{.80}
\plotone{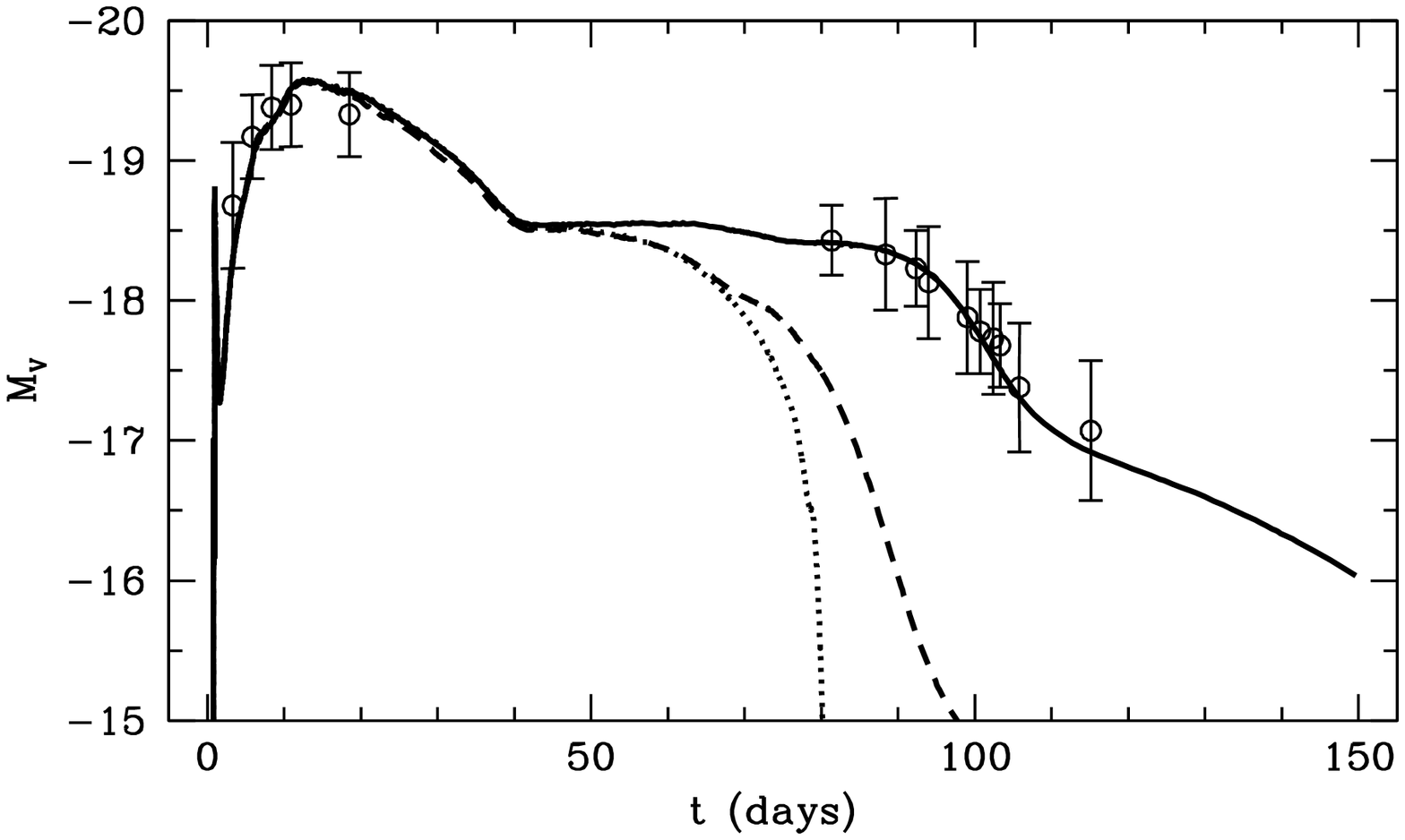}
\caption{The calculated V light curve of the optimal model (\emph{solid line\/})
         is compared with the observations of SN 2009kf (\citet{Bot_10})
         (\emph{open circles\/}).
         The effect of the adopted $^{56}$Ni mass is illustrated for the cases of
         the $^{56}$Ni mass equal to $0.0765~M_{\sun}$ (\emph{dashed line\/})
         and the zero $^{56}$Ni mass (\emph{dotted line\/}).
         \label{fig1}}
\end{figure}

\begin{figure}
\epsscale{.80}
\plotone{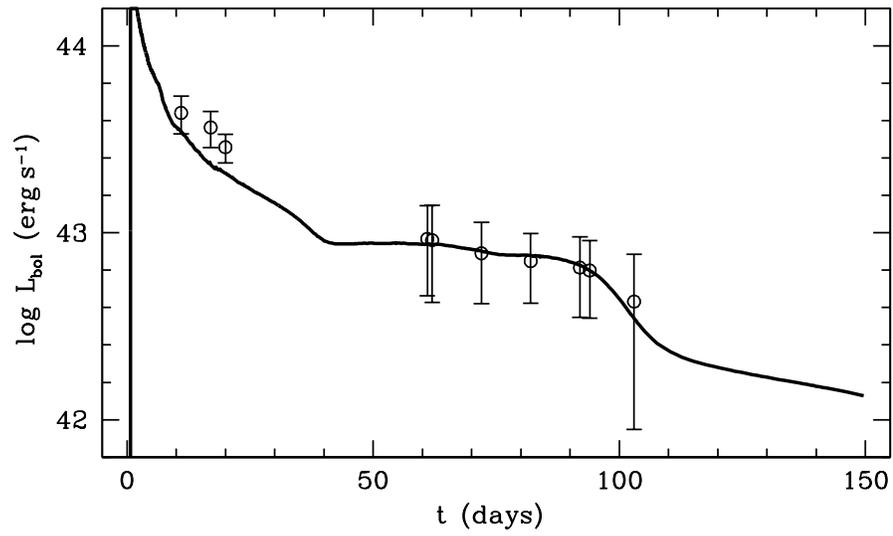}
\caption{The calculated bolometric light curve of the optimal model
      (\emph{solid line\/}) overplotted on the bolometric data of the black-body fit
      for SN 2009kf (\emph{open circles\/}).
\label{fig2}}
\end{figure}

\begin{figure}
\epsscale{.80}
\plotone{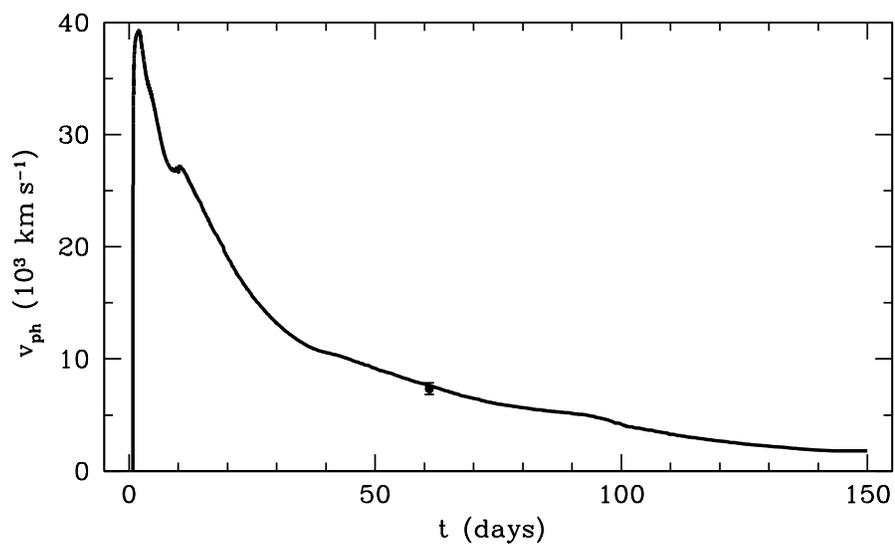}
\caption{The calculated photospheric velocity (\emph{solid line\/}) in comparison
         with the velocity at the photosphere recovered from the H$\alpha$ and
         He\,{\sc i} 5876 \AA\ line profiles on day 61 (\emph{filled circle\/}).
      \label{fig3}}
\end{figure}

The extensive study of a parameter space led us to the optimal model of SN~2009kf that
   can describe simultaneously the $V$ light curve (Fig. \ref{fig1}), the early strong
   bolometric luminosity peak along with the end phase of the plateau (Fig. \ref{fig2}),
   and the expansion velocity at the photosphere (Fig. \ref{fig3}).
The observational $V$ light curve is taken from \citet{Bot_10}, and the bolometric light
   curve is constructed with a black-body fit to the $g$, $r$, $i$, $z$ photometry and
   the NUV flux, only for the first three epochs, assuming a total extinction $A_V=1$ mag.
The velocity at the photosphere of $7350\pm350$ km\,s$^{-1}$ is recovered from the spectrum
   on day 61 \citep{Bot_10} using the modelling of the H$\alpha$ and He\,{\sc i} 5876 \AA\
   line profiles.
In fitting the light curves we rely primarily on the $V$ band data rather than on the
   bolometric light curve in which the initial peak is mainly determined by the NUV flux
   which in turn is severely hampered by errors in the total extinction.

The parameter set of the optimal model consists of the ejected mass
   $M_{\rm env}=28.1~M_{\sun}$, explosion energy $E=2.2\times10^{52}$ erg,
   pre-SN radius $R_0=2\times10^3~R_{\sun}$, and the $^{56}$Ni mass of $0.4~M_{\sun}$.
The latter is the upper limit reported by \citet{Bot_10}.
On the basis of the $V$ light curve and the velocity at the photosphere we determine
   the SN parameters with accuracy of $\sim$ 10\%.
The ejecta mass combined with the adopted mass of the black hole of $4.5~M_{\sun}$ gives
   a pre-SN mass of $32.6\pm3~M_{\sun}$, maximal among other pre-SNe for previously
   modelled SNe~IIP \citep[cf.][]{UC_09}.

It should be emphasized that for the adopted value of the $^{56}$Ni mass the ejecta mass
   and the explosion energy of the model are minimal; a decrease of the $^{56}$Ni mass
   would require a larger ejecta mass and explosion energy.
The $V$ light curve for the model with the $^{56}$Ni mass of $0.0765~M_{\sun}$ and
   the model without $^{56}$Ni illustrate this point (Fig. \ref{fig1}): with the lower
   $^{56}$Ni mass the plateau is shorter, so a larger ejecta mass and explosion energy
   are needed to account for the plateau duration for the same photospheric velocity.
Computations show that in order to produce sensible fit of the bolometric light curve
   with $0.0765~M_{\sun}$ of $^{56}$Ni one needs to increase the ejected mass and
   explosion energy by a factor of $\sim 1.5$.
The light curve at the end of the plateau depends also on the extent of the $^{56}$Ni
   mixing.
Remarkably, in the model with the $^{56}$Ni mass of $0.4~M_{\sun}$ the $^{56}$Ni should
   be mixed homogeneously in the inner ejecta up to 7700 km\,s$^{-1}$.

The luminous broad initial peak of the bolometric light curve of SN~2009kf is related,
   as usually for SNe~IIP, with the large pre-SN radius.
The recovered radius $R_0\sim2\times10^3~R_{\sun}$ is larger than the radii of pre-SNe
   of well-studied SNe~IIP, which lie in the range of $35-1500~R_{\sun}$
   (Table~\ref{tbl-1}).
Keeping in mind that the radius of massive RSG increases with the progenitor mass
   \citep{HJLB_97}, the large pre-SN radius can be considered as an independent evidence
   in favor of the relatively large pre-SN mass compared to normal SNe~IIP.

The primary uncertainty of the derived SN parameters is related with the poorly
   determined extinction in the host galaxy.
We explore this uncertainty assuming a total extinction to be $A_V=0$ and $A_V=1.5$ mag.
In the case of $A_V=0$ mag the upper limit of $^{56}$Ni mass is $0.16~M_{\sun}$.
With this amount of $^{56}$Ni the pre-SN radius should be smaller, $R_0=200~R_{\sun}$,
   while the ejecta mass and the explosion energy of the model increase up to
   $\sim 40~M_{\sun}$ and $\sim 3\times10^{52}$ erg, respectively.
Note that the $E/M$ ratio is to be invariant because the velocity at the photosphere
   on day 61 is fixed.
For $A_V=1.5$ mag the upper limit of the $^{56}$Ni mass is $0.63~M_{\sun}$.
This amount of $^{56}$Ni permits us to obtain a satisfactory fit of the light curves
   to the observations with the same ejecta mass and the explosion energy as for
   the optimal model, but with the pre-SN radius increased by a factor of $\sim$ 2.
However, if the adopted $^{56}$Ni mass is lower than the upper limit, then one needs
   to increase the ejecta mass and the explosion energy to reproduce the light curves
   together with the velocity at the photosphere on day 61.
We thus conclude that the ejecta mass and explosion energy cannot be lower than the
   values for the optimal hydrodynamic model.

\section{Discussion}
\label{sec-disc}

\begin{figure}
\epsscale{.80}
\plotone{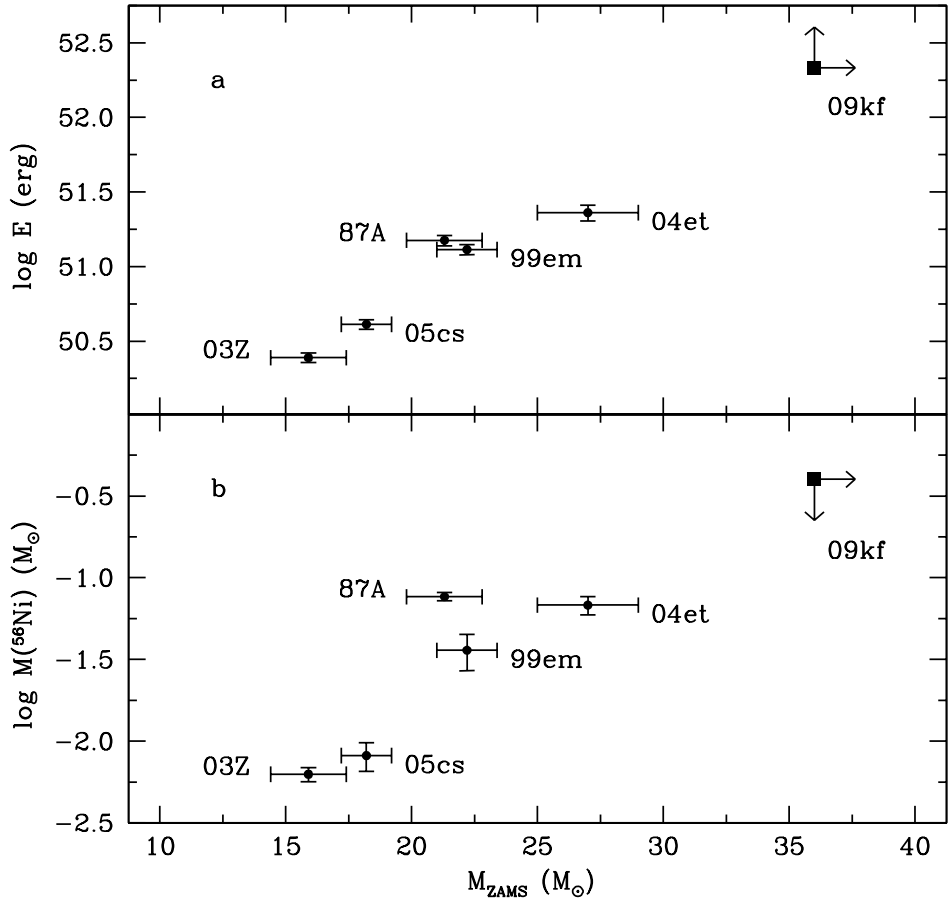}
\caption{Explosion energy \textbf{a}) and $^{56}$Ni mass \textbf{b}) versus
      hydrodynamic progenitor mass for six type IIP supernovae.
\label{fig4}}
\end{figure}

The major result of the hydrodynamic modelling of SN~2009kf is the conclusion that
   the SN has been caused by the energetic explosion of a massive and extended RSG.
Parameters of hydrodynamic model for SN~2009kf are listed in Table~\ref{tbl-1} together
   with parameters of other SNe~IIP studied earlier \citep[see][]{UC_09}.
The columns in the order contain SN name, pre-SN radius, ejecta mass, explosion energy,
   $^{56}$Ni mass, maximal velocity of the $^{56}$Ni mixing zone, minimal velocity
   of the hydrogen matter, mass of the collapsed core, pre-SN mass, mass lost by
   the wind, and progenitor mass.
In the case of SN~2009kf progenitor we suggest a single star scenario, while the total
   mass lost through the wind outflow is adopted to be equal to that lost by the
   progenitor of SN~2004et.
SN~2009kf stands out from the list of other SNe~IIP by its extremely large explosion
   energy which is 10 times greater than the explosion energy of SN~2004et, the most
   energetic event among previously studied SNe~IIP.
An exceptional nature of SN~2009kf is emphasized by its position on diagrams of
   the explosion energy and the $^{56}$Ni mass vs. progenitor mass (Fig. \ref{fig4}).
Despite the fact that SN~2009kf does not deviate from the general trends on both plots
   the explosion physics of SN~2009kf is likely essentially different from that of other
   SNe~IIP.

Before the further discussion of implications of the results we should consider
   alternative explanations for the unusually high luminosity of SN~2009kf.
Given the rather confident upper limit for the $^{56}$Ni amount, the only remaining
   possibility for the power source is the CS interaction. 
This mechanism suggests that the kinetic luminosity, released in the radiative shock wave,
   is emitted by the cool dense shell (CDS) at the ejecta/CS interface likewise it is
   the case for SNe~IIn.
Quick look at the spectra of SN~2009kf at the plateau stage shows that the strong CS
   interaction is unlikely in this case, because the spectra have little to do with
   the SN~IIn spectra.
Indeed, the latter never show pronounced broad absorption lines \citep[cf.][]{Fil_97};
   by contrast, the H$\alpha$ line of SN~2009kf has the deep absorption component
   \citep[see][]{Bot_10}, quite similar to other SNe~IIP.
To strengthen further arguments against the CS interaction we note that the maximal
   velocity detected in the H$\alpha$ absorption component of SN~2009kf on day 61 is
   $\sim13000$ km\,s$^{-1}$.
This means that the double-shock structure, formed by the forward and reverse shocks with
   the CDS at the contact surface, expands with the velocity $>10^4$ km\,s$^{-1}$.
Meanwhile, the H$\alpha$ absorption component indicates that the continuum source resides
   at velocities $v<6000$ km\,s$^{-1}$, which is in odd with the assumption that the bulk
   radiation originates from the high-velocity ejecta/CS interface.
This evidence thus rules out the CS interaction as a major source of the SN~2009kf
   luminosity.

Regardless of the role of the CS interaction, the upper limit of the wind density could be
   valuable.
The wind density can be estimated using the upper limit of the luminosity at the nebular
   stage $M_R>-16.3$ mag on day 236 \citep{Bot_10}, which can be converted into the upper
   limit of the optical luminosity, $L<5\times10^{41}$ erg\,s$^{-1}$, assuming that
   the spectral energy distribution is similar to that at the end of the plateau stage.
To recover the upper limit of the wind density we employ the interaction model based on
   the thin shell approximation \citep{CCU_07} for the ejecta mass and the explosion
   energy found for SN~2009kf.
For the adopted $^{56}$Ni mass of $0.4~M_{\sun}$ the upper limit of the wind density
   parameter turns out to be $w=\dot{M}/u=6\times10^{15}$ g\,cm$^{-1}$.
This estimate refers to the epoch of 236 days and the radius of $6\times10^{16}$ cm.
Assuming the wind velocity $u=10$ km\,s$^{-1}$ we conclude that the upper limit of the
   wind density corresponds to the mass-loss rate
   $\dot{M}<9\times10^{-5}~M_{\sun}$\,yr$^{-1}$
   at the stage $\sim2\times10^3$ yr before the SN explosion.

The question arises on the nature of the central "infernal machine" responsible for the
   huge explosion energy $E\approx2\times10^{52}$ erg, one order magnitude exceeding the
   energy of normal SNe~IIP.
In fact, the explosion energy of SN~2009kf is comparable with the explosion energy of
   energetic SNe~Ic (hypernovae), associated with gamma-ray bursts (e.g. SN~1998bw).
According to the recent compilation of \citet{Tan_09}, the explosion energies of
   hypernovae lie in the range of $(0.6-5)\times10^{52}$ erg; SN~2009kf falls into this
   interval.
The absence of alternative possibilities compels us to suggest that SN~2009kf is caused
   by the "engine" associated with the hypernova phenomenon.
The possibility that the collapsar mechanism could operate in massive stars with retained
   hydrogen envelope and produce a very powerful SN was proposed by \citet{MWH_01} and
   \citet{WHW_02}.

The widely shared view is that the hypernova explosion is powered by
   a rapid disk accretion into the black hole \citep{MW_99,MWH_01,HFWLH_03}.
Two conceivable scenarios are proposed how the black hole and accretion disk system
   with the high accretion rate might arise: (i) a core collapse in a single rapidly
   rotating star (collapsar model) \citep{Woo_93,MW_99}, or (ii) a black hole
   (neutron star) merger with the He-core of massive companion (merger model)
   \citep{FW_98}.
The merger scenario suggests that the neutron star, merged with He-core as a result of
   a super-Eddington accretion, rapidly grows a black hole surrounded by the accretion
   disk.

The presence of a massive hydrogen envelope in SN~2009kf imposes constraint on the single
   star scenario: the progenitor should be less massive than the critical mass $M_{WR}$
   above which WR stars form due to the complete removal of the hydrogen envelope.
This boundary is not well known because of uncertainties in the mass-loss rate;
   for the solar metallicity the currently preferred value is in the range of
   $30-40~M_{\sun}$ \citep{HFWLH_03}.
Thus, in the single star scenario the progenitor mass should be $<40~M_{\sun}$.

The merger scenario includes two major stages: (i) conservative evolution of a close
   massive binary that ends up with the formation of the neutron star (black hole)
   in pair with the massive normal star, and (ii) spiral-in of the neutron star
   (black hole) in the massive star companion with subsequent merging with the helium
   core.
The neutron star in the process of merging grows black hole \citep{Che_93} so that
   the final stage of pre-SN is the central black hole surrounded with the dense helium
   accretion disk embedded in the massive hydrogen envelope.

As an example, the original close binary might consist of $M_1=25~M_{\sun}$ and
   $M_2=20~M_{\sun}$ stars.
After the conservative mass transfer the primary ends up as $5-8~M_{\sun}$ He star
   in pair with the secondary $37-40~M_{\sun}$ star.
The subsequent evolution of the helium star leads to SN~Ib/c or SN~IIb explosion with
   the neutron star (black hole) left behind in pair with the massive star.
This is the stage of massive X-ray binary.
After the secondary forms a He core, the stellar radius grows, the Roche lobe overflow
   begins and the common envelope forms.
The common envelope stage ends up as a black hole surrounded with the helium accretion
   disk within the massive hydrogen envelope.
The rapid disk accretion onto black hole eventually gives rise to the SN~2009kf event.

The energetic SNe~IIP produced by the hypernova mechanism are extremely rare events.
This is indicated by the large distance of the unique SN~2009kf.
To a first approximation the rate ratio of SN~2009kf-like to SN~1998bw-like events
   can be estimated interpreting their distances ($D_1=740$ Mpc and $D_2=40$ Mpc,
   respectively) as the closest neighbor distances.
The closest neighbor distance $R$ is defined by the obvious equation
   $(1/3) R^3 \Omega \eta G t = 1$,
   where $\Omega$ is the solid angle of the survey, $\eta$ is the detection efficiency,
   $G$ is the SN rate per unit volume, and $t$ is the total observation time.
Assuming that $\Omega$, $\eta$, and $t$ are comparable for the detection of both
   hypernovae and SN~2009kf-like events, we obtain the rough estimate of the rate ratio
   $G_1/G_2 \sim (D_2/D_1)^3\sim10^{-4}$.
Note that Malmquist bias cannot change this ratio significantly because the absolute
   magnitudes of these phenomena are comparable.
We conclude therefore that roughly 0.01\% of all hypernovae can retain a massive hydrogen
   envelope and produce SN~2009kf-like events.

\acknowledgments

Research of V. P. U. has been supported in part by RFBR under grant 10-02-00249-a.
M. T. B. acknowledges support from the award "Understanding the lives of
   massive stars from birth to supernovae" (S.J. Smartt) made under
   European Heads of Research Councils and European Science Foundation
   EURYI Awards scheme.

\clearpage

\clearpage

\clearpage

\clearpage

\clearpage

\begin{deluxetable}{ l c c c c c c c c c c c }
\tabletypesize{\scriptsize}
\tablecaption{Hydrodynamic models of type IIP supernovae and their progenitor masses
   \label{tbl-1}}
\tablewidth{0pt}
\tablehead{
\colhead{SN} &\colhead{$R_0$} &\colhead{$M_{env}$} &\colhead{$E$} &\colhead{$M_{\mathrm{Ni}}$}
   &\colhead{$v_{\mathrm{Ni}}^{max}$} &\colhead{$v_{\mathrm{H}}^{min}$}
   &\colhead{$M_{NS}$} &\colhead{$M_{\rm pre-SN}$} &\colhead{$M_{lost}$}
   &\colhead{$M_{\rm ZAMS}$} &\colhead{ref} \\
\colhead{} &\colhead{$(R_{\sun})$} &\colhead{$(M_{\sun})$} &\colhead{($10^{51}$ erg)}
   &\colhead{$(10^{-2} M_{\sun})$} &\colhead{(km\,s$^{-1}$)} &\colhead{(km\,s$^{-1}$)}
   &\colhead{$(M_{\sun})$} &\colhead{$(M_{\sun})$} &\colhead{$(M_{\sun})$}
   &\colhead{$(M_{\sun})$} &\colhead{}
}
\startdata
 1987A &  35  & 18   & 1.5   & 7.65 &  3000 & 600 & 1.6 & 19.6 & 1.7 & 19.8--22.8 & 1 \\
1999em & 500  & 19   & 1.3   & 3.60 &  660  & 700 & 1.6 & 20.6 & 1.6 & 21.0--23.4 & 2 \\
 2003Z & 229  & 14   & 0.245 & 0.63 &  535  & 360 & 1.4 & 15.4 & 0.2--0.8 & 14.4--17.4 & 3 \\
2004et & 1500 & 22.9 & 2.3   & 6.8  &  1000 & 300 & 1.6 & 24.5 & 1.4--3.4 & 25.0--29.0 & 4 \\
2005cs & 600  & 15.9 & 0.41  & 0.82 &  610  & 300 & 1.4 & 17.3 & 1.0 & 17.6--20.4 & 5 \\
2009kf & 2000 & 28.1 & 21.5  & 40.0 &  7700 & 410 & 4.5\tablenotemark{a} & 32.6 & 3.4 &
       $\ge36$\tablenotemark{b} & 6 \\
\enddata
\tablecomments{Table \ref{tbl-1} gives mass estimates of type IIP progenitors according to
   the following relation: $M_{\rm ZAMS}$ = $M_{NS}$ or $M_{BH}$ + $M_{env}$ + $M_{lost}$.}
\tablenotetext{a}{It is the mass of black hole}
\tablenotetext{b}{In a single star scenario}
\tablerefs{
(1) Utrobin 2004;
(2) Utrobin 2007;
(3) Utrobin, Chugai, \& Pastorello 2007;
(4) Utrobin \& Chugai 2009;
(5) Utrobin \& Chugai 2008;
(6) the present work.
}
\end{deluxetable}

\end{document}